\documentclass{aa}
\usepackage{graphicx}

\def\jh{\mbox{$\rm (J-H)$}}

\def\mMo{\mbox{$\rm (m-M)_O$}}
\def\ebv{\mbox{$\rm E(B-V)$}}
\def\ejh{\mbox{$\rm E(J-H)$}}
\def\rc{\mbox{$\rm R_{core}$}}
\def\rl{\mbox{$\rm R_{lim}$}}
\def\rext{\mbox{$\rm R_{extr}$}}
\def\ms{\mbox{$\rm M_\odot$}}
\def\ds{\mbox{$\rm d_\odot$}}
\def\dgc{\mbox{$\rm d_{GC}$}}
\def\jj{\mbox{$\rm J$}}
\def\hh{\mbox{$\rm H$}}
\def\ks{\mbox{$\rm K_S$}}
\def\mevol{\mbox{$\rm m_{evol}$}}
\def\mobs{\mbox{$\rm m_{obs}$}}
\def\mtot{\mbox{$\rm m_{tot}$}}
\def\kms{\mbox{$\rm km\,s^{-1}$}}
\def\tr{\mbox{$\rm t_{relax}$}}
\def\tcr{\mbox{$\rm t_{cross}$}}

\begin{document}

\title{Properties of five low-contrast open clusters in the third quadrant}

\author{E. Bica\inst{1} \and C. Bonatto\inst{1}}

\offprints{Ch. Bonatto}

\institute{Universidade Federal do Rio Grande do Sul, Instituto de F\'\i sica, 
CP\,15051, Porto Alegre 91501-970, RS, Brazil\\
\mail{charles@if.ufrgs.br}
}

\date{Received --; accepted --}

\abstract{We derive photometric, structural and dynamical evolution-related 
parameters of five as yet unstudied low-contrast open clusters located in the third quadrant 
using 2MASS data. The target clusters are Czernik\,31, Czernik\,32, Haffner\,9, Haffner\,11 
and Trumpler\,13. We apply a statistical field-star decontamination procedure to infer on 
the intrinsic colour-magnitude diagram (CMD) morphology which is critical for such low-contrast 
objects. Consequently, it became possible to derive accurate reddening, age, distance from the 
Sun and Galactocentric distance for the five clusters. In the structural and luminosity/mass-function 
analyses we apply a colour-magnitude filter which encompasses the cluster evolutionary CMD 
sequences and excludes stars with  discrepant colours. Using this procedure we derive core and 
limiting radii, mass function (MF) slope, total mass, mass density and relaxation time. We 
derive ages in the range 140 -- 1\,100\,Myr, Galactocentric distances within 7.7 -- 11.4\,kpc, 
and total masses within 360--2\,900\,\ms. Reflecting large-scale mass segregation, the MF slope 
in the core is significantly flatter than that in the halo of the five clusters. Although some of 
the present clusters are relatively younger than the Gyr-old clusters, they present evidence 
of advanced dynamical evolution. This kind of study has become possible because of the photometric 
uniformity and spatial coverage of 2MASS which allows a proper subtraction of the field-star 
contamination on the target CMDs. The present study indicates that low-contrast clusters can be 
studied with 2MASS, particularly after field-star subtraction, which is important since most of 
the unstudied open clusters belong to this class.

\keywords{({\it Galaxy}:) open clusters and associations: general; 
{\it Galaxy}: structure} }

\titlerunning{Low-contrast open clusters}

\authorrunning{E. Bica \and C. Bonatto}

\maketitle

\section{Introduction}
\label{intro}

Open clusters are formed in and are distributed throughout the  Galactic disk. Subsequent interactions
with the disk associated with the relentless tidal pull of the Galactic center/bulge drive their 
dynamical evolution and tend to destroy the less-massive ones in a time-scale of a few 
$\rm10^8\,yr$ (Bergond, Leon, \& Guibert \cite{Bergond2001}). Those surviving to older ages may
reach greater vertical distances into the thick disk. Consequently, photometric, structural and 
dynamical parameters of open clusters turn out to be excellent probes of the Galactic structure, 
star-formation processes and evolution (e.g. Lyng\aa\ \cite{Lynga82}; Janes \& Adler \cite{JA82}; 
Friel \cite{Friel95}). The dynamical evolution of open clusters has been investigated using N-body 
codes (e.g. de la Fuente Marcos \cite{delaF98}) and through the determination of a set of 
observational parameters (Bonatto, Bica, \& Santos Jr. \cite{BBS2005}; Bonatto \& Bica 
\cite{BB2005}).

In Galactic structure studies it is fundamental to have as complete a census of the open clusters
as possible, since their spatial distribution can be used to better constrain theoretical mass 
models of the Galaxy. However, since Janes \& Adler (\cite{JA82}) the number 
of open clusters with derived parameters  (reddening, age and distance from the Sun) increased 
from 434 to 631, taking as reference the clusters 
currently with parameters in the WEBDA\footnote{\em http://obswww.unige.ch/webda} open cluster database 
(Mermilliod \cite{Merm1996}). In the recent revision of open clusters Dias et al. (\cite{Dias2002}) 
report a total of 1537 clusters catalogued. There is a need to explore this majority of unstudied 
clusters.

In recent years the 2MASS\footnote{The Two Micron All Sky Survey, All Sky data release (Skrutskie et 
al. \cite{2mass1997}), available at {\em http://www.ipac.caltech.edu/2mass/releases/allsky/}} database 
has proven to be a fundamental tool in the analysis of open clusters with different brightness and contrast.  
The 2MASS Point Source Catalogue (PSC) is uniform, reaching relatively faint magnitudes and covering nearly 
all the sky, allowing a proper field-star definition both for low-contrast clusters and those with 
large angular sizes. We mention the discovery and analysis of two faint open clusters in Cygnus (Bica, 
Bonatto, \& Dutra \cite{BBD2003}) and three in other parts of the sky (Bonatto, Bica \& Dutra 
\cite{BBD2004}) using 2MASS photometry. In Bonatto \& Bica (\cite{BB2005}) we used 2MASS to obtain a 
homogeneous set of parameters, compare dynamical states and derive mass functions (MFs) for 11 
relatively populous, nearby open clusters.

Considering the above we decided to study five open clusters in the third quadrant using 2MASS
data and the techniques outlined in Bonatto \& Bica (\cite{BB2005}). Because the present sample is
composed of low-contrast clusters (Sects.~\ref{2mass} and \ref{struc}) we include in the analysis 
an algorithm to statistically decontaminate the observed CMDs of Galaxy (field) stars to better  
define the CMD morphology. This step is essential to unambiguously separate 
low-contrast open clusters from fluctuations in field-star counts. In addition, colour-magnitude 
filters encompassing the cluster evolutionary CMD sequences are subsequently used in the structural 
and luminosity/mass-function analyses.
 
For the present clusters we derive the age, reddening, distance from the Sun, Galactocentric distance,
core and limiting radii, mass and density, MF slope and relaxation time. We derive parameters for 
the cluster as a whole as well as the core and halo subsystems. Besides the interest in obtaining 
parameters of as yet unstudied clusters, this paper represents a test of the limits of 2MASS (and 
our techniques) when dealing with more distant, low-contrast, low-Galactic latitude clusters.

Uncertainties throughout this paper correspond to $\rm1\sigma$ Poisson statistics.

This paper is organized as follows. In Sect.~\ref{OCS} we present the sample and show XDSS R 
images of the clusters. In Sect.~\ref{2mass} we present the 2MASS photometric parameters and
introduce the field-star decontamination procedure. In Sect.~\ref{struc} we analyze the spatial 
structure of the clusters. In Sect.~\ref{MF} we derive the mass functions and discuss stellar 
content properties. In Sect.~\ref{comp} we compare the present clusters with nearby, populous 
open clusters which span a range of dynamical states. Concluding remarks are given in 
Sect.~\ref{Conclu}. 

\section{The third quadrant (3Q) open cluster sample}
\label{OCS}

The clusters selected for the present study are Czernik\,31 (OCl-625, ESO\,560SC\,3),
Czernik\,32 (Ki\,24, OCl-683, BH\,11, ESO\,494SC\,20), Haffner\,9 (OCl-600), Haffner\,11 
(OCl-657, BH\,3, ESO\,429SC\,3, and Trumpler\,13 (Cr\,219, OCl-815, BH\,94, ESO\,127SC\,17). 
Additional designations above are from Alter, Ruprecht \& Vanisek (\cite{Alter1970}), Lauberts 
(\cite{lau82}), and van den Bergh \& Hagen (\cite{BH75}). In what follows we will use the acronyms 
Cz, Haf and Tr to refer to the clusters. For simplicity we will refer to the present clusters as 
the 3Q sample.

\begin{figure*}
\begin{minipage}[b]{0.33333\linewidth}
\includegraphics[width=\textwidth]{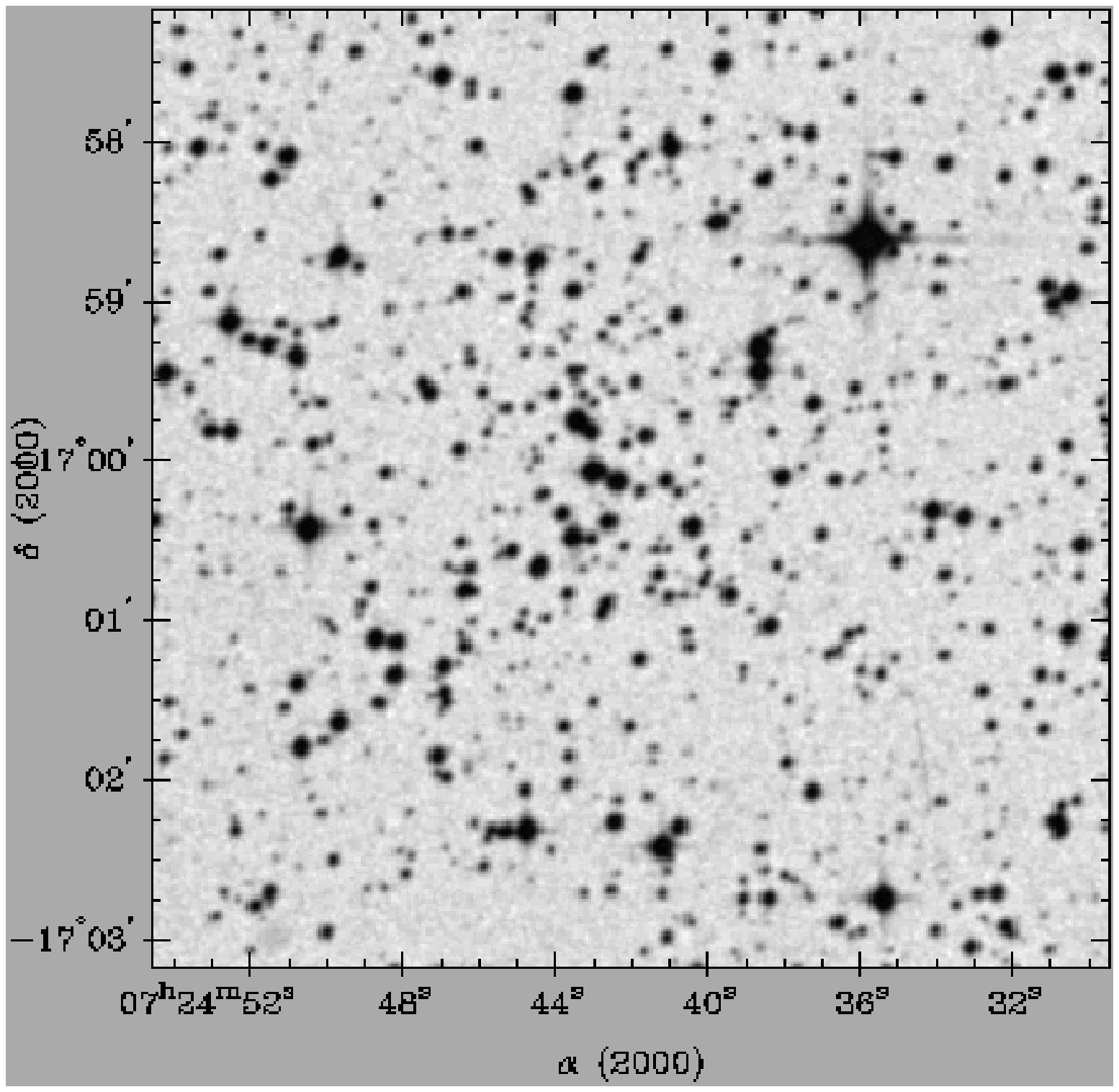}
\end{minipage}\hfill
\begin{minipage}[b]{0.33333\linewidth}
\includegraphics[width=\textwidth]{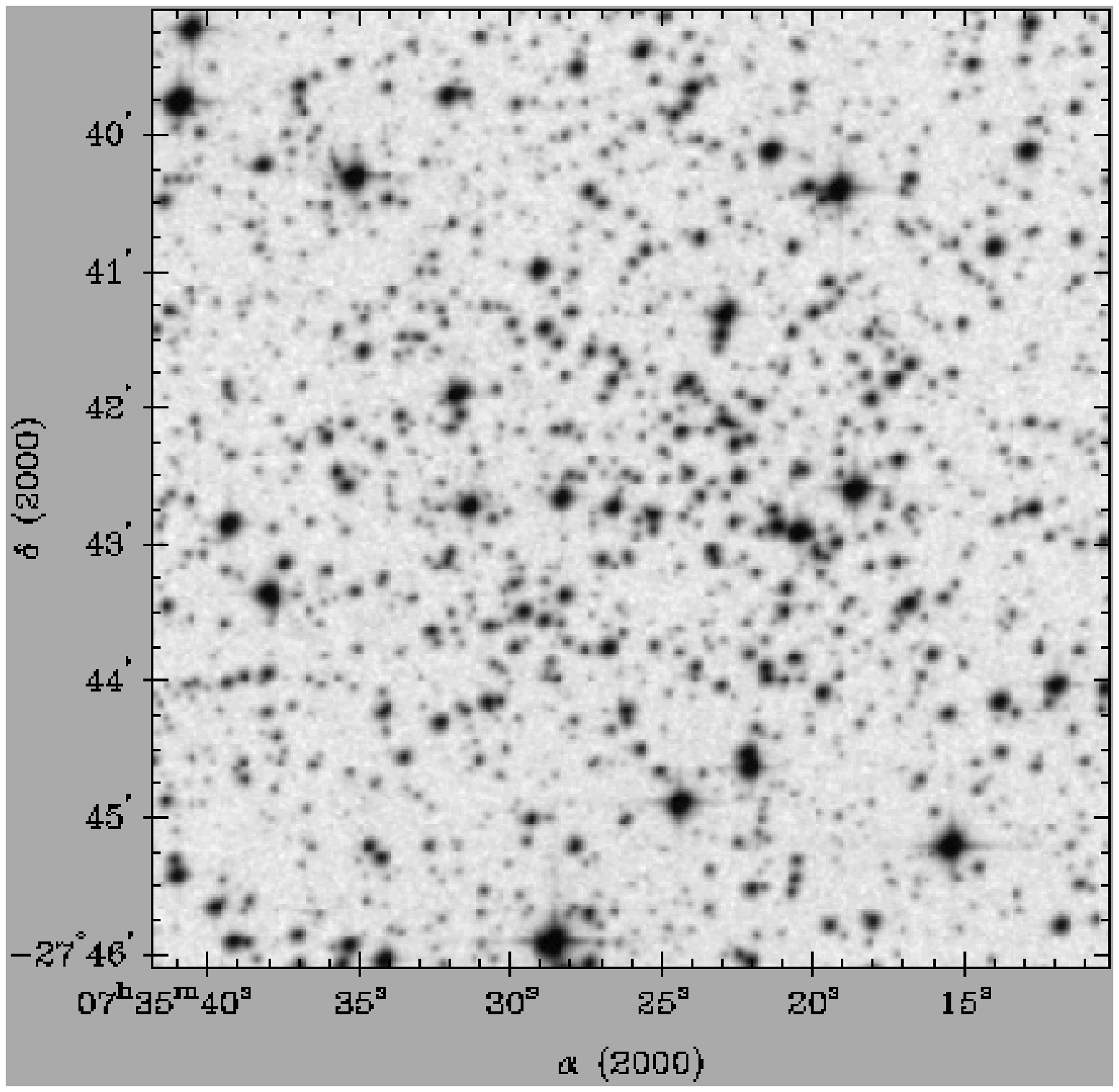}
\end{minipage}\hfill
\begin{minipage}[b]{0.33333\linewidth}
\includegraphics[width=\textwidth]{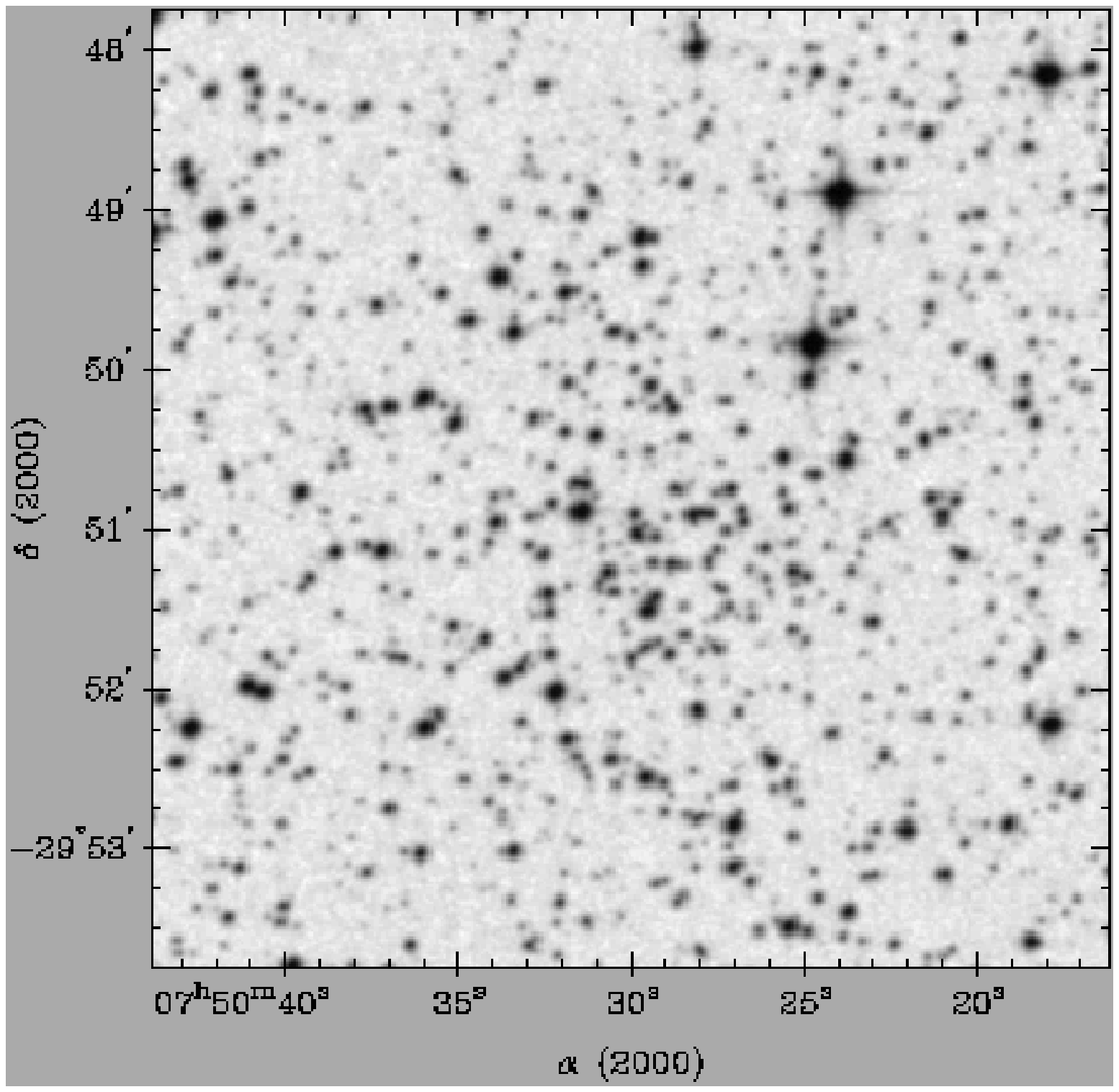}
\end{minipage}\hfill
\caption[]{XDSS R images. Left panel: Haf\,9 ($6\arcmin\times6\arcmin$). Middle panel: Haf\,11 
($7\arcmin\times7\arcmin$). Right panel: Cz\,32 ($6\arcmin\times6\arcmin$).}
\label{fig1}
\end{figure*}

\begin{figure*}
\begin{minipage}[b]{0.33333\linewidth}
\includegraphics[width=\textwidth]{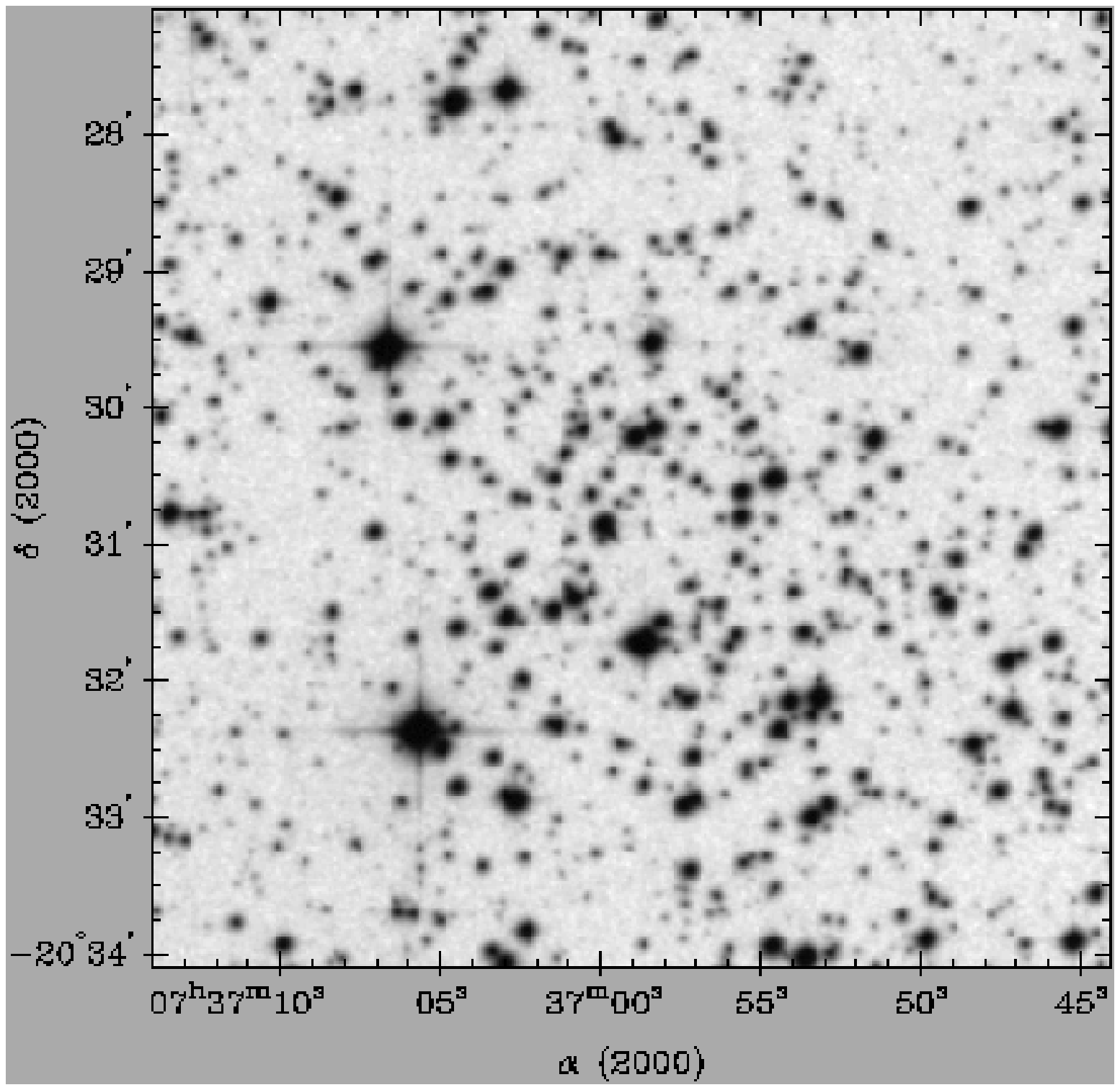}
\end{minipage}\hfill
\begin{minipage}[b]{0.33333\linewidth}
\includegraphics[width=\textwidth]{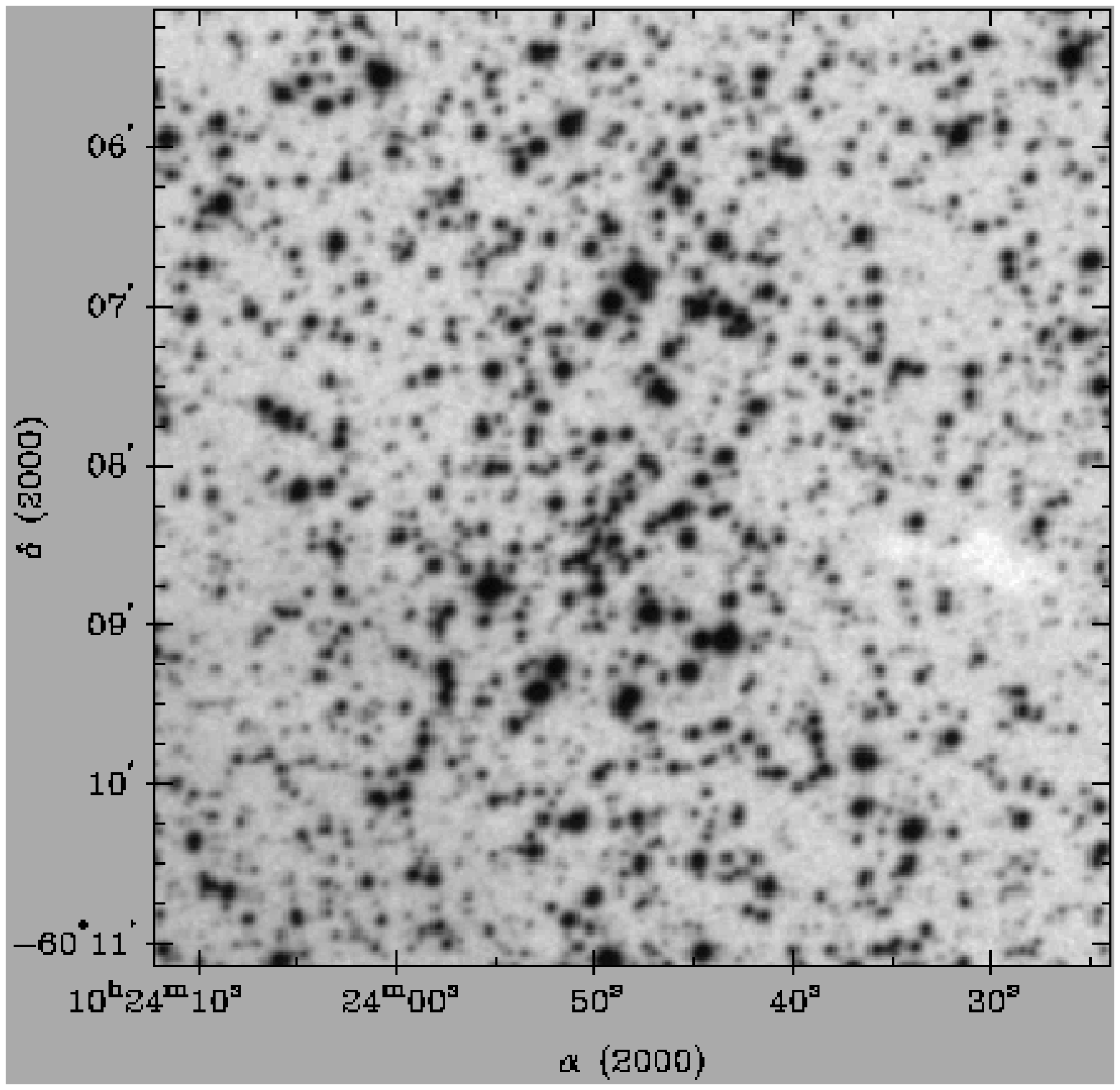}
\end{minipage}\hfill
\caption[]{XDSS R images. Left panel: Cz\,31 ($7\arcmin\times7\arcmin$). Right panel: 
Tr\,13 ($6\arcmin\times6\arcmin$).}
\label{fig2}
\end{figure*}

In Figs.~\ref{fig1} and \ref{fig2} we present XDSS R images of the clusters in the 3Q sample.
DSS and XDSS images can be extracted from the Canadian Astronomy Data Centre (CADC\footnote{\em 
http://cadcwww.dao.nrc.ca/}). In all panels a central concentration of stars with varying 
star-density contrast with respect to the background can be seen. 

Accurate central coordinates are fundamental for the photometric, and primarily structural, 
analysis of open clusters. This is particularly critical in the derivation of the core radius 
and mass of small clusters (see e.g. Bonatto \& Bica \cite{BB2005}). For the 3Q sample clusters
we indicate in Table~\ref{tab1} different sources of coordinates. The original Alter, Ruprecht
\& Vanisek (\cite{Alter1970}) coordinates (1st line for each cluster in Table~\ref{tab1}) are 
not accurate enough for the present purposes of deriving core, halo and overall parameters. 
Consequently we measured in the DSS and XDSS images more precise optical centers for each cluster. 
Figs.~\ref{fig1} and \ref{fig2} are centered on these coordinates.

\section{The 2MASS photometry and near-infrared CMDs}
\label{2mass}

The VizieR\footnote{\em http://vizier.u-strasbg.fr/viz-bin/VizieR?-source=II/246} tool was 
used to obtain \jj, \hh\ and \ks\ 2MASS photometry. The central coordinates corresponding
to the density peaks in the 2MASS photometry of each cluster were derived by examining histograms 
of the number of stars in 0.5\arcmin\ bins of right ascension and declination. In what follows we 
refer to the coordinates that maximize the density of stars (Sect.~\ref{struc}) at $\rm r=0\arcmin$ 
as the cluster center. These central coordinates are given in columns~2 and 3 of Table~\ref{tab1}. In  
a few cases the central coordinates are slightly shifted with respect to the optical centers 
(Figs.~\ref{fig1} and \ref{fig2}). For each cluster we extracted photometry of the stars contained 
in a circular area with radius \rext\ (column~7 of Table~\ref{tab1}) centered on the respective central 
coordinates. 

\begin{figure} 
\resizebox{\hsize}{!}{\includegraphics{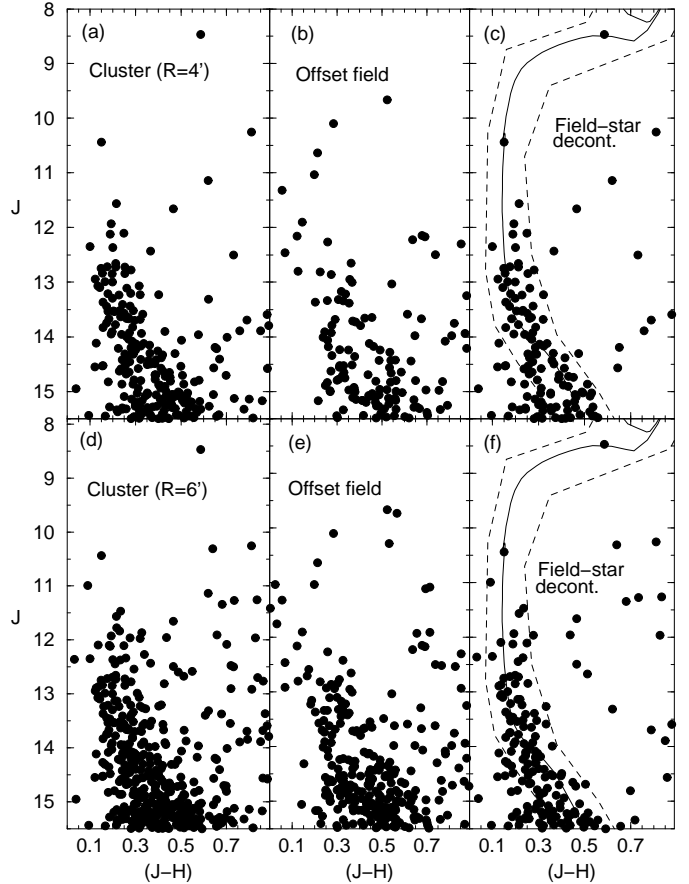}}
\caption[]{$\jj\times\jh$ CMDs of the central 4\arcmin\ extraction of Haf\,9 (panel (a)) and
respective same-area offset field (panel (b)). The corresponding field-star decontaminated CMD 
is in panel (c). Bottom panels: same as above for the 6\arcmin\ extraction. Solid line: 
{\em best-fit} Padova isochrone with $\rm age=140\,Myr$. Dashed line: colour-magnitude filter
(Sect.~\ref{struc}) used in the radial and luminosity/mass-function analyses.}
\label{fig3}
\end{figure}

\begin{figure} 
\resizebox{\hsize}{!}{\includegraphics{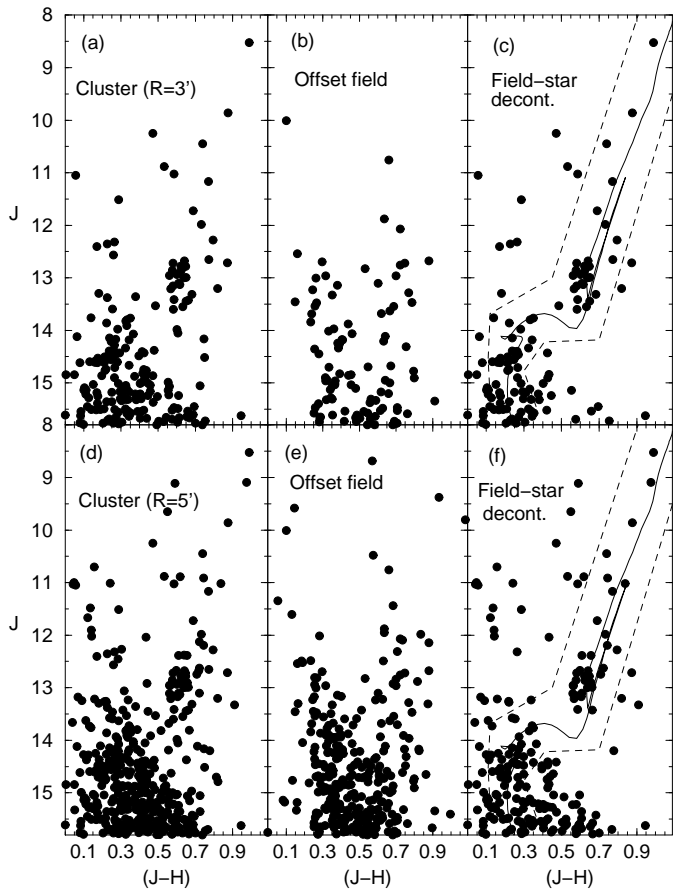}}
\caption[]{Same as Fig.~\ref{fig3} for Haf\,11. The Padova isochrone corresponds to the
age 890\,Myr. Stars above the turnoff may be blue-stragglers.}
\label{fig4}
\end{figure}

\begin{figure} 
\resizebox{\hsize}{!}{\includegraphics{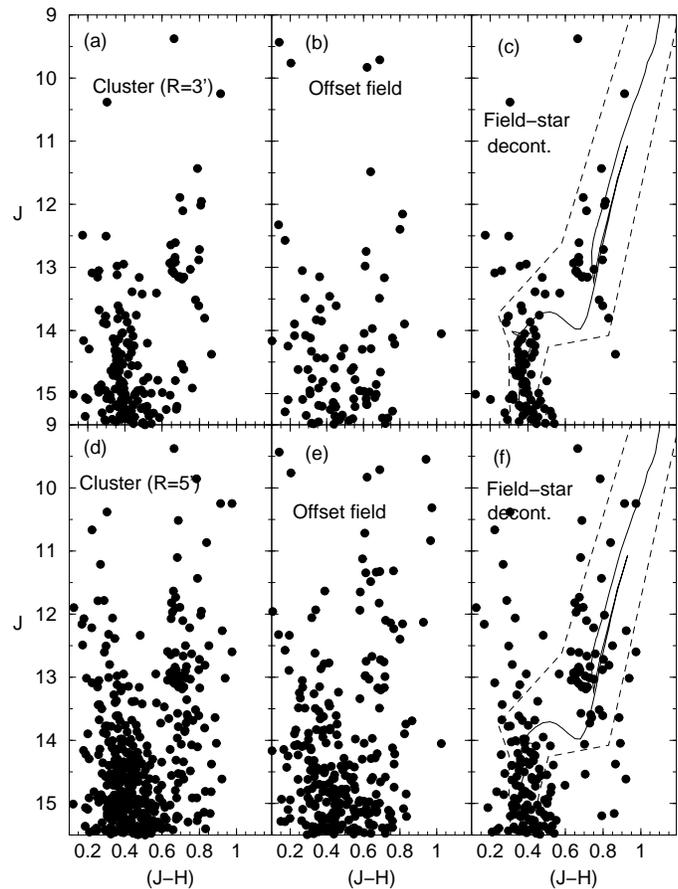}}
\caption[]{Same as Fig.~\ref{fig3} for Cz\,32. The Padova isochrone corresponds to the
age 1.1\,Gyr. Stars above the turnoff may be blue-stragglers.}
\label{fig5}
\end{figure}

\begin{figure} 
\resizebox{\hsize}{!}{\includegraphics{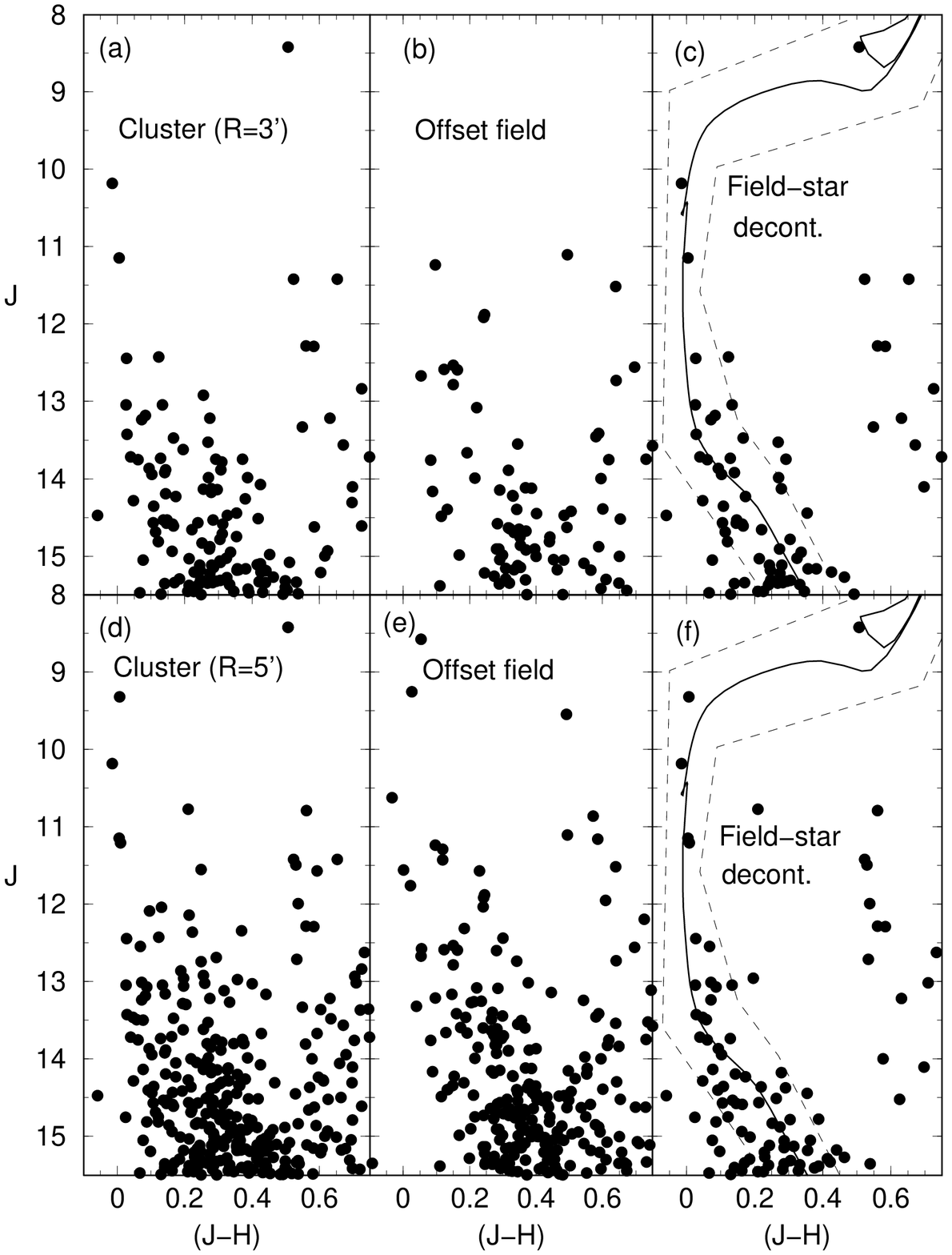}}
\caption[]{Same as Fig.~\ref{fig3} for Cz\,31. The Padova isochrone corresponds to the
age 180\,Myr.}
\label{fig6}
\end{figure}

\begin{figure} 
\resizebox{\hsize}{!}{\includegraphics{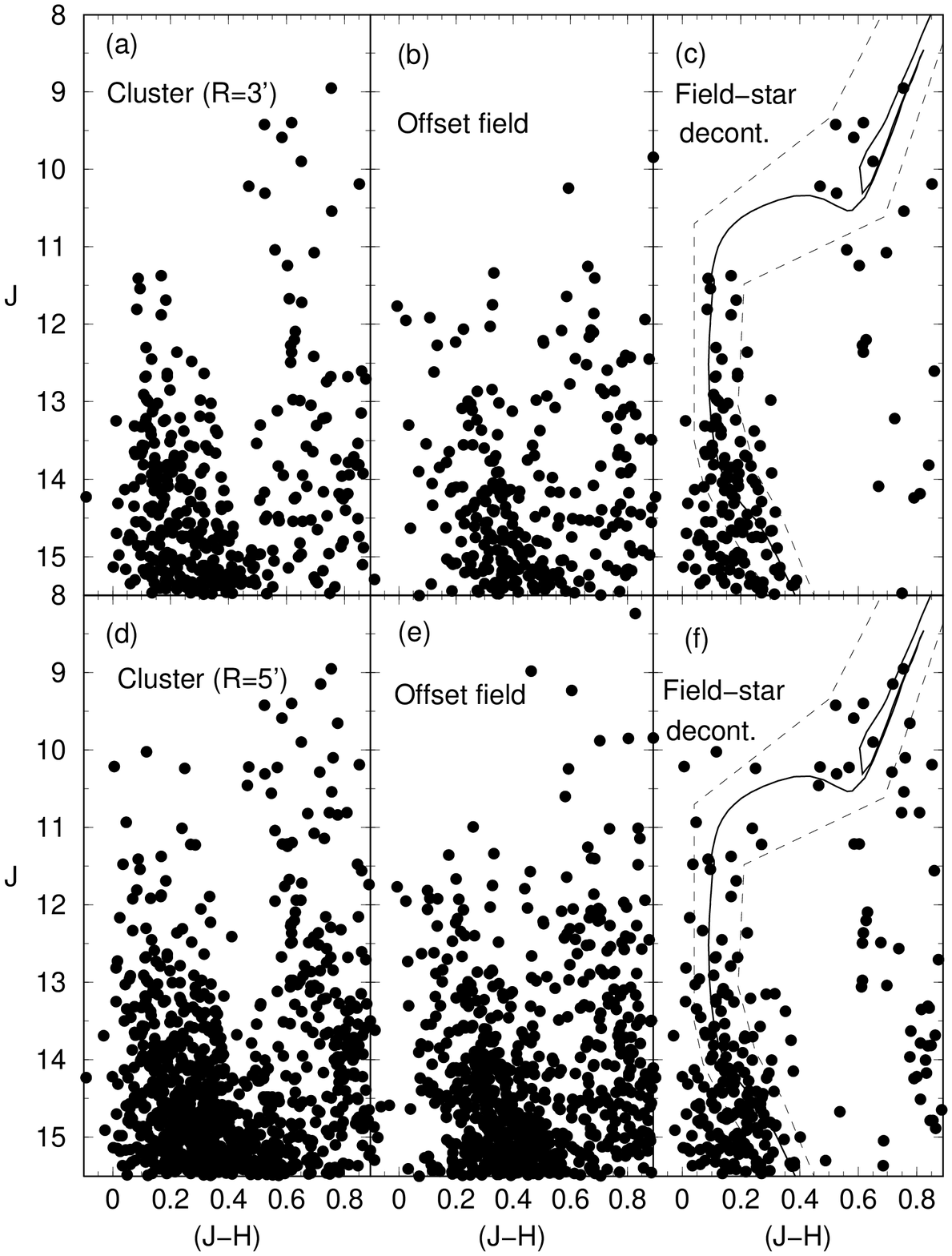}}
\caption[]{Same as Fig.~\ref{fig3} for Tr\,13. The Padova isochrone corresponds to the
age 320\,Myr.}
\label{fig7}
\end{figure}

Because of the relatively low Galactic latitudes of the 3Q sample clusters (column~5 of Table~\ref{tab1}), 
the target fields present significant field-star contamination, mostly from disk stars. 
The relative amount of this contamination with respect to the central field turns out to be 
important in the identification and analysis of small and/or poor open clusters. To illustrate
the relative contribution of the field stars with respect to varying spatial areas we show in 
the left panels of Figs.~\ref{fig3} to \ref{fig7} the $\jj\times\jh$ CMDs of the target clusters 
in two different extractions. The corresponding (same area) offset fields are shown in the middle 
panels. The smaller extraction (panels (a)) is that maximizing the CMD density contrast with 
respect to the offset field. The smaller radius corresponds to about $\rm4\times$ the core radius 
(\rc, Sect.~\ref{struc}). The radius of the larger extraction (panels (b)) is intermediate between 
that of the smaller extraction and the limiting radius (\rl, Sect.~\ref{struc}). The offset fields 
were built with the stars in rings more external than the limiting radius.

The cluster CMDs contrast in morphology and density with respect to the offset fields, although
field-star contamination in some cases is important, e.g. Cz\,31 and Cz\,32. To minimize the 
visual effect of the field stars on the cluster CMDs we applied a decontamination procedure 
which considers two different approaches for field-star subtraction: {\em (i)} absolute counts 
and {\em (ii)} nearest-neighbour density counts. In case {\em (i)} the algorithm divides the CMD 
in boxes of dimensions $\rm\Delta\jj=0.75$ and $\rm\Delta\jh=0.075$, both for the cluster and 
offset field. The effective number of cluster stars in a given CMD box corresponds to the absolute 
difference in the number of cluster and offset-field stars. After this step the algorithm randomly 
excludes from each cluster CMD box the required number of stars; boxes resulting with a negative 
number of stars are left blank. In case {\em (ii)} we calculate the star density in CMD boxes with 
dimensions $\rm\Delta\jj=0.35$ and $\rm\Delta\jh=0.035$. The observed star density in a given box 
corresponds to the average number of stars in it and the 1st, 2nd and 3rd-neighbouring row and 
columns, both for the cluster and offset-field regions. Subtraction of the observed offset-field 
density from that of the cluster yields the effective cluster star density in each CMD box. 
Multiplying the ratio of the effective/observed star density by the observed number of stars 
yields the effective number of cluster stars in a given CMD box. From this point on the algorithm
proceeds as in approach {\em (i)} to exclude stars in each CMD box. Comparison of the resulting
decontaminated CMDs with a variety of offset fields indicated that the best results were obtained 
with approach {\em (ii)}, probably because it is less sensitive to field-star count fluctuations 
than approach {\em (i)}. To maximize the statistical representativity of the field-star counts we 
consider as offset field the region $\rm 1.5\times\rl\leq r\leq R_{ext}$. The ratio between the 
cluster and offset-field spatial areas is considered when we calculate the respective number (or 
density) of stars. Approach {\em (i)} is similar to that used by Kerber \& Santiago (\cite{KS05}), 
and {\em (ii)} is similar to Mighell et al. (\cite{MRSF96}).

The resulting field-star decontaminated CMDs are shown in the right panels of Figs.~\ref{fig3} to 
\ref{fig7}. The decontaminated CMDs are cleaner than the observed ones (left panels) and 
present morphologies typical of open clusters of different ages. In particular the giant clumps of 
Haf\,11 (Fig.~\ref{fig4}) and Cz\,32 (Fig.~\ref{fig5}) show up in the respective CMDs which in turn 
helps constrain their ages and distances from the Sun. The field-star decontaminated CMDs are used 
to better define the intrinsic CMD morphology and isochrone fit. 

\begin{table*}
\caption[]{3Q cluster coordinates and 2MASS photometric parameters}
\begin{scriptsize}
\label{tab1}
\renewcommand{\tabcolsep}{0.80mm}
\renewcommand{\arraystretch}{1.5}
\begin{tabular}{ccccccccccccccc}
\hline\hline
Object & $\rm\alpha(J2000)$ & $\rm\delta(J2000)$ & $\ell$ & $b$&z & \rext &$\delta^\prime_c$&Age & \ebv & \mMo &$\rm M_J$&\ds &Scale &\dgc \\
       &  (h:m:s) & $(^\circ:\arcmin:\arcsec)$ & $(^\circ)$ & $(^\circ)$&(pc) & (\arcmin)& & (Myr) &(mag)&(mag)&(mag)& (kpc)&(pc/\arcmin) & (kpc)\\
(1)&(2)&(3)&(4)&(5)&(6)&(7)&(8)&(9)&(10)&(11)&(12)&(13)&(14)&(15)\\
\hline
Haf\,9$^\dag$&07:24:30&-17:00\\
Haf\,9       &07:24:42.0&-17:00:10.0&$231.80$&$-0.59$&$-20\pm2$&30&$3.9\pm0.9$&$140\pm20$&$0.54\pm0.03$&$11.4\pm0.1$&3.1&$1.9\pm0.2$&1.78&$9.3\pm0.2$\\

Haf\,11$^\dag$&07:35:24&-27:44\\
Haf\,11       &07:35:30.0&-27:41:58.6&$242.39$&$-3.52$&$-322\pm15$&40&$2.8\pm0.5$&$890\pm150$&$0.36\pm0.03$&$13.6\pm0.1$&1.1&$5.2\pm0.2$&0.66&$11.4\pm0.2$\\

Cz\,32$^\dag$&07:50:24&-29:50 \\
Cz\,32       &07:50:30.0&-29:50:45.0&$245.89$&$-1.73$&$-121\pm8$&30&$4.7\pm1.2$&$1120\pm200$&$0.66\pm0.03$&$13.0\pm0.1$&1.4&$4.0\pm0.2$&0.85&$10.3\pm0.3$\\

Cz\,31$^\dag$&07:36:48&-20:30 \\
Cz\,31       &07:36:59.0&-20:30:35.0&$236.27$&$+0.27$&$+11\pm1$&50&$3.4\pm1.2$&$180\pm20$&$0.06\pm0.03$&$11.7\pm0.1$&3.2&$2.2\pm0.2$&1.53&$9.4\pm0.2$\\

Tr\,13$^\dag$ &10:23:48&-60:05 \\
Tr\,13        &10:23:49.2&-60:07:32.5&$285.51$&$-2.34$&$-100\pm6$&15&$1.9\pm0.4$&$320\pm40$&$0.30\pm0.03$&$11.9\pm0.1$&2.8&$2.4\pm0.2$&1.41&$7.7\pm0.2$\\

\hline\hline
\end{tabular}
\begin{list}{Table Notes.}
\item Col.~6: distance to the Galactic plane. Col.~7: Extraction radius of the 2MASS photometry.  
Col.~8: Observed contrast parameter $\rm\delta^\prime_c$. Col.~9: {\em Best-fit} isochrone age.
Col.~10: Colour-excess resulting from the isochrone fit.
Col.~11: intrinsic distance modulus. Col.~12: faint-absolute magnitude limit of the stars in the 
radial-density profiles. Col.~13: distance from the Sun. Col.~14: parsec to arcmin scale. Col.~15: 
Galactocentric distance. $(\dag)$: coordinates from Alter et al. (\cite{Alter1970}).
\end{list}
\end{scriptsize}
\end{table*}

Photometric parameters were derived by means of solar metallicity Padova isochrones (Girardi et al. 
\cite{Girardi2002}) computed with the 2MASS \jj, \hh\ and \ks\ filters\footnote{\em 
http://pleiadi.pd.astro.it/isoc\_photsys.01/isoc\_photsys.01.html}, fitted to the 
field-star decontaminated CMDs of the central cluster regions in Figs.~\ref{fig3} to \ref{fig7} 
(right panels). The 2MASS transmission filters produced isochrones very similar to the Johnson ones, 
with differences of at most 0.01 in \jh\ (Bonatto, Bica \& Girardi \cite{BBG2004}). For reddening and 
absorption transformations we use $\rm R_V=3.2$, and the relations $\rm A_J=0.276\times A_V$ and 
$\ejh=0.33\times\ebv$, according to Dutra, Santiago \& Bica (\cite{DSB2002}, and references therein). 
2MASS photometric uncertainties as a function of magnitude are discussed in Soares \& Bica 
(\cite{SB2002}) and Bonatto, Bica \& Santos Jr. (\cite{BBS2005}).

Taking into account the uncertainties associated with the isochrone fit we derive for each cluster in
the 3Q sample the age (column~9 in Table~\ref{tab1}) colour excess \ejh\ and the corresponding \ebv\
(column~10), the absolute distance modulus \mMo\ (column~11) and the distance from the Sun \ds\ 
(column~13). The Galactocentric distance (column~15) was calculated using 8.0\,kpc as the distance 
of the Sun to the center of the Galaxy (Reid \cite{Reid93}). The resulting isochrone fits are shown 
in the field-star decontaminated CMDs (right panels of Figs.~\ref{fig3} to \ref{fig7}). 

3Q cluster ages are $\rm140\pm20\,Myr$ (Haf\,9), $\rm180\pm20\,Myr$ (Cz\,31), $\rm320\pm40\,Myr$ 
(Tr\,13), $\rm890\pm150\,Myr$ (Haf\,11) and $\rm1.1\pm0.2\,Gyr$ (Cz\,31). Except for Cz\,31, the 
remaining clusters in the 3Q sample are considerably reddened with $\rm 0.30\leq\ebv\leq0.66$. They 
are distant from the Sun with $\rm 1.9\leq\ds(kpc)\leq5.2$ and, except for Tr\,13, the remaining 3Q 
clusters are located outside the Solar circle.

The relatively low-contrast nature of the clusters in the 3Q sample is apparent in the XDSS R images 
(Figs.~\ref{fig1} and \ref{fig2}) and is reflected in the low values of the observed contrast 
parameter $\delta^\prime_c=1.9-4.7$ (column~8 of Table~\ref{tab1}). $\delta^\prime_c$ represents 
the ratio of the number of stars in the cluster region with respect to that (same area) in the 
offset field. Additional clues to their nature as open clusters are {\em (i)} the cluster-like 
field-star decontaminated CMD morphology and isochrone fits (Figs.~\ref{fig3} to \ref{fig7}), 
{\em (ii)} the King-like radial-density profiles (Sect.~\ref{struc}), and {\em (iii)} the 
resulting MFs (Sect.~\ref{MF}).

\section{Cluster structure}
\label{struc}

Structural parameters of the 3Q clusters were derived by means of the radial density profile, i.e. 
the projected number of stars per area around the cluster center. Before counting stars we applied 
the colour-magnitude filters (dashed line in the right panels of Figs.~\ref{fig3} to \ref{fig7}) to the 
observed CMDs to discard stars with  discrepant colours (mostly Galactic field stars) and maximize 
cluster membership. This filtering procedure was previously applied in the analysis of the open 
clusters M\,67 (Bonatto \& Bica \cite{BB2003}), NGC\,188 (Bonatto, Bica \& Santos Jr. \cite{BBS2005}) 
and NGC\,3680 (Bonatto, Bica \& Pavani \cite{BBP2004}). This procedure minimizes the 
probability of field-star contamination, spurious detections and the increase of photometric 
uncertainties at faint magnitudes. We give in column~12 of Table~\ref{tab1} the resulting absolute 
faint-magnitude ($\rm M_J$) limit of the stars included in each radial-density profile.

The radial density profiles were obtained by counting stars inside concentric annuli with steps of 
1.0\arcmin\ (Cz\,31 and Haf\,11) and 0.5\arcmin\ (Cz\,32, Haf\,9 and Tr\,13) in radius. In each case 
the field-star contribution level corresponds to the average number of stars included in the outermost 
ring. 

\begin{figure} 
\resizebox{\hsize}{!}{\includegraphics{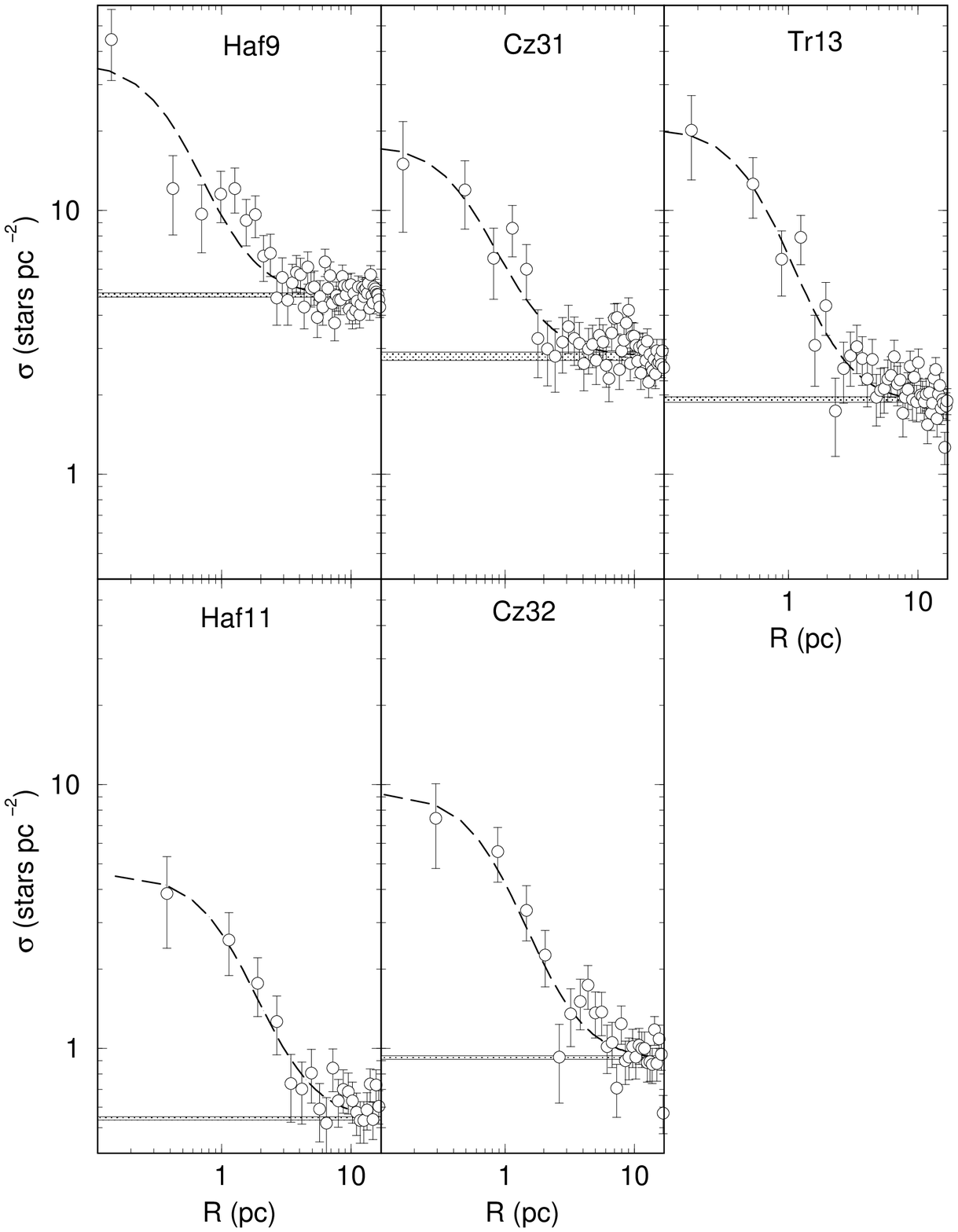}}
\caption[]{Radial density profiles of the 3Q clusters. The respective two-parameter King model fits 
are shown. To allow absolute comparison between clusters, all profiles are plotted in the same scale 
with parsecs in the abscissa and $\rm stars\,pc^{-2}$ in the ordinate.}
\label{fig8}
\end{figure}

The resulting radial density profiles are shown 
in Fig.~\ref{fig8}. For absolute comparison between clusters we scale the radius in the abscissa 
in parsecs, and the number density of stars in the ordinate in $\rm stars\,pc^{-2}$\ using the 
distances derived in Sect.~\ref{2mass}. The statistical significance of each profile is reflected 
in the $1\sigma$\ Poisson error bars. The cluster limiting radius (\rl) can be estimated 
considering the fluctuations in the radial density profile with respect to the field stars. In
this sense, \rl\ corresponds to the region where the density profile merges with the background. 
For regions beyond $\rl$ the null-contrast between cluster and field-star density would produce 
prohibitive Poisson errors and meaningless results. For practical purposes the bulk of the cluster 
stars are contained within $\rl$. The tidal radius derived from the three-parameter King (\cite{King1962}) 
profile can only be obtained for much more populated open clusters, such as e.g. NGC\,188, M\,67, 
NGC\,2477 and M\,26 (Bonatto \& Bica \cite{BB2005}).

Cluster structural parameters were derived by fitting the two-parameter King (\cite{King1966a}) 
surface density profile to the background-subtracted radial distribution of stars. The two-parameter 
King model essentially describes the intermediate and central regions of normal clusters (King 
\cite{King1966b}; Trager, King \& Djorgovski \cite{TKD95}). The fit was performed using a nonlinear 
least-squares fit routine that uses the errors as weights. The best-fit solutions are shown in 
Fig.~\ref{fig8} superimposed on the respective radial density profiles. The King profile provides
a good representation of the radial distribution of stars in all clusters in the 3Q sample, which 
reinforces their open cluster nature. Since it follows from an isothermal (virialized) sphere, the 
close similarity of the radial distribution of stars with a King profile suggests that the internal 
structure of the clusters (particularly the core) have already reached some significant level 
of dynamical evolution. We will return to this point in Sects.~\ref{dyna} and \ref{comp}. The 
structural parameters are given in Table~\ref{tab2}. 

\begin{table}
\caption[]{3Q cluster structural parameters}
\begin{scriptsize}
\label{tab2}
\renewcommand{\tabcolsep}{1.03mm}
\renewcommand{\arraystretch}{1.25}
\begin{tabular}{cccccc}
\hline\hline
Object&$\rm\sigma_{bg}$ & $\sigma_{0K}$& $\delta_c$ &\rc & \rl \\
&$\rm (stars\, pc^{-2})$ & $\rm (stars\, pc^{-2})$ &  &(pc) & (pc)\\
(1)&(2)&(3)&(4)&(5)&(6)\\
\hline
Haf\,9 &$4.79\pm0.08$&$32.2\pm9.2$&$7.8\pm1.9$&$0.42\pm0.10$&$4.7\pm0.6$\\
Haf\,11&$0.54\pm0.01$&$4.0\pm1.7$&$8.6\pm3.3$&$1.09\pm0.33$&$12.1\pm1.5$\\
Cz\,32 &$0.55\pm0.01$&$8.4\pm3.5$&$10.4\pm3.9$&$0.81\pm0.23$&$7.8\pm0.6$\\
Cz\,31 &$2.75\pm0.01$&$15.1\pm6.5$&$6.4\pm2.3$&$0.53\pm0.18$&$4.0\pm0.3$\\
Tr\,13 &$1.92\pm0.05$&$19.0\pm7.7$&$11.2\pm4.2$&$0.59\pm0.17$&$7.1\pm0.7$\\
\hline\hline
\end{tabular}
\begin{list}{Table Notes.}
\item The background stellar density (col.~2) was measured in the outermost ring.
The central density (col.~3) and core radius (col.~5) were derived from the King fit, 
$\rm\sigma(r)=\sigma_{bg}+\sigma_{0K}/(1+(r/R_{core})^2)$. Col.~4: density contrast 
parameter $\rm\delta_c=\sigma(0)/\sigma_{bg}=1+\sigma_{0K}/\sigma_{bg}$. The limiting 
radius (col.~6) was estimated directly from the density profiles. 
\end{list}
\end{scriptsize}
\end{table}

We quantify the cluster/background star-density contrast by means of the parameter $\delta_c$ 
(column~4), defined as the ratio of the central density ($\sigma(0)$) to that of the background 
($\rm\sigma_{bg}$). The $\delta_c$ values in Table~\ref{tab2} can be compared to those of the 
rather populous and high-Galactic latitude open clusters (Bonatto \& Bica \cite{BB2005}) NGC\,188 
($b=+22.39^\circ$; $\ds\sim1.7$\,kpc; $\rm M\sim3800\,\ms$) with $\delta_c=21.5\pm2.9$, and M\,67 
($b=+31.89^\circ$; $\ds\sim0.9$\,kpc; $\rm M\sim990\,\ms$) with $\delta_c=33.7\pm5.7$, and that 
of the populous and rather low-latitude cluster NGC\,2477 ($b=-5.82^\circ$; $\ds\sim1.2$\,kpc; 
$\rm M\sim5300\,\ms$) with $\delta_c=19.7\pm1.9$. The  determination of the density contrast 
parameter seems to depend more on the Galactic latitude than cluster mass. Because the 3Q sample 
is composed of low-latitude 
clusters their contrast parameters are significantly lower than those of the comparison 
clusters but higher than the observed ones ($\rm\delta^\prime_c$, col.~8 of Table~\ref{tab1}).
As a consequence, some underestimation of the limiting radii can be expected in most  of the clusters
studied in the present paper.
Deeper wide-field photometry would be important to check this possibility.

\section{Mass functions}
\label{MF}

Based on the King profile fits we decided to derive MFs $\left(\phi(m)=\frac{dN}{dm}\right)$ for 
the core, halo and overall regions. As offset field we consider the region from $\approx1.5\times\rl$ 
to $\rm R_{ext}$ (Table~\ref{tab1}), which provides statistical representativity in terms of background 
distribution of stellar luminosities (and mass), as shown in Bonatto, Bica \& Santos Jr. 
(\cite{BBS2005}) and Bonatto \& Bica (\cite{BB2005}).

\begin{figure} 
\resizebox{\hsize}{!}{\includegraphics{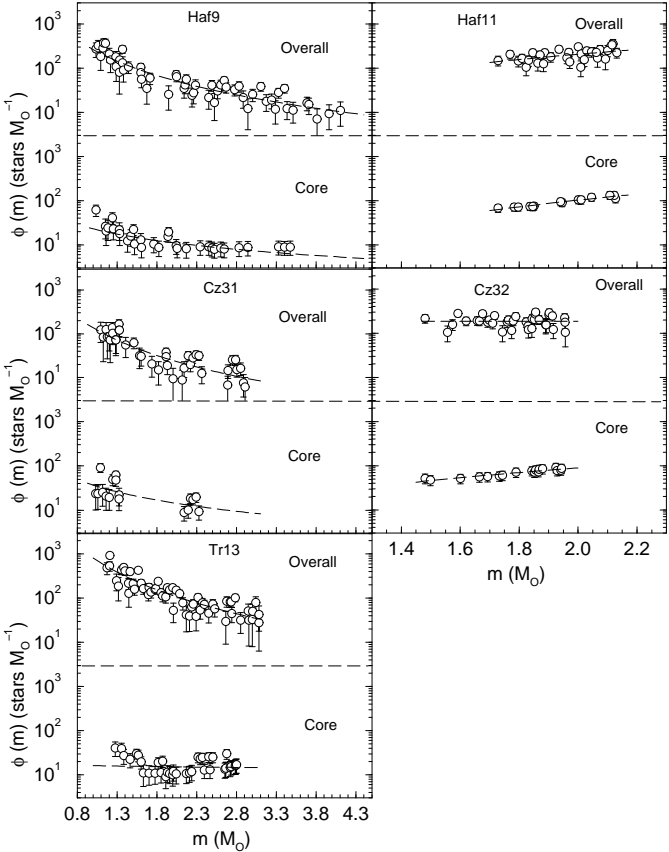}}
\caption[]{Core and overall mass functions of the 3Q clusters. Each panel contains MFs derived from 
the \jj, \hh\ and \ks\ 2MASS photometry. MF fits $\left(\phi(m)\propto m^{-(1+\chi)}\right)$ are shown 
as dashed lines, and the respective MF slopes are given. In all cases the core MF is flatter than
the overall MF.}
\label{fig9}
\end{figure}

In the case of low-latitude clusters it is essential that the Galactic field contamination of the 
CMDs is properly taken into account in order to derive the intrinsic luminosity and mass distributions 
of the member stars. To do this we first apply the colour-magnitude filter (right panels of 
Figs.~\ref{fig3} to \ref{fig7}) to both cluster and offset field CMDs. The filtering process takes 
into account most of the background, leaving a residual contamination. We deal with this residual 
contamination statistically by building the luminosity functions (LFs) for each cluster region and 
offset field. The three 2MASS bands are treated independently, taking into account the 99.9\% PSC 
Completeness Limit\footnote{Corresponding to the Level\,1 Requirement, according to 
{\em\tiny http://www.ipac.caltech.edu/2mass/releases/allsky/doc/sec6\_5a1.html}}. Consequently, 
the faint magnitude limit of each LF is $\jj=15.8$, $\hh=15.1$\ and $\ks=14.3$, respectively. We take 
the turnoff as the bright limit to avoid inconsistencies in the mass-luminosity relation. For each 
2MASS band we build a LF by counting stars in magnitude bins from the respective faint magnitude limit 
to the turnoff, both for each cluster region and offset field. Considering that the solid angle of the 
offset field is different from that of a given cluster region, we multiply the offset field LF by a 
numerical factor so that the solid angles match. The intrinsic LF of each cluster region is obtained 
by subtracting the respective (i.e. solid angle-corrected) offset-field LF from that of the cluster
region. Finally, the intrinsic LFs are transformed into MFs using the mass-luminosity relation obtained 
from the respective Padova isochrone and distance modulus (Sect.~\ref{2mass}). These procedures 
are repeated independently for the three 2MASS bands (Bonatto \& Bica \cite{BB2005}). The final MF of 
a given cluster region is produced by combining the \jj, \hh\ and \ks\ MFs into a single MF. The 
resulting core and overall MFs of the 3Q sample clusters are shown in Fig.~\ref{fig9}.

We provide in Table~\ref{tab3} parameters derived from the LFs and MFs for the  core ($\rm R\leq\rc$), 
halo ($\rm\rc\leq R\leq\rl$) and overall ($\rm core+halo$) regions. The number of evolved stars (column~3) 
is calculated by integrating the intrinsic 
LFs for magnitudes brighter than the turnoff. Multiplying this by the mass at the turnoff gives an 
estimate of the evolved-star mass (column~4). This procedure produces a realistic value of the number 
of member evolved stars because the background contamination was statistically subtracted from the 
LF. The observed main sequence (MS) mass range is in column~5. The MF slope of the MS stars is in 
column~6. The core MFs are significantly flatter than halo MFs (see also Sect.~\ref{dyna}). The 
number of MS stars and corresponding mass are derived by integrating the MF from the faint magnitude 
limit to the turnoff. We add to these the corresponding values of the number and mass of evolved 
stars to derive the total number of observed stars (column~7) and observed mass (column~8). 

The large distances from the Sun of the clusters in the 3Q sample preclude detection of sub-Solar 
mass stars. However, an estimate of the total mass locked up in MS stars can be made by taking into 
account all stars from the turnoff down to the H-burning mass limit, $0.08\,\ms$, assuming the universal 
IMF of Kroupa (\cite{Kroupa2001}), in which $\chi=0.3\pm0.5$ for the range $\rm 0.08\leq m(\ms)\leq0.5$ 
and $\chi=1.3\pm0.3$ for $\rm 0.5\leq m(\ms)\leq1.0$. For the cases where the present value of $\chi$ 
is smaller or of the same order than Kroupa's we extrapolate our MFs. The resulting extrapolated values 
of the number of stars and extrapolated mass (added to the corresponding values for the evolved stars) 
are given respectively in cols.~9 and 10 of Table~\ref{tab3}. We also give in cols.~11 and 12 the 
projected and volume mass densities, respectively. 

\begin{table*}
\caption[]{Measurements and parameters derived from the MFs.}
\label{tab3}
\renewcommand{\tabcolsep}{0.95mm}
\begin{tabular}{ccccccccccccccc}
\hline\hline
&&\multicolumn{2}{c}{Evolved}&&\multicolumn{2}{c}{MS}&&\multicolumn{2}{c}{Observed+Evolved}&
&\multicolumn{4}{c}{Extrapolated+Evolved}\\
\cline{3-4}\cline{6-7}\cline{9-10}\cline{12-15}\\
OC&Region&N$^*$&\mevol&&$\Delta m$ &$\chi_{MS}$&&$\rm N^*$&\mobs&&$\rm N^*$&\mtot&$\sigma$&$\rho$\\
&&(Stars)&(\ms)&&(\ms)& &&($10^2$stars)&($10^2\ms$)&& ($10^2$stars)&
($10^2\ms$)&($\rm \ms\,pc^{-2}$)&($\rm \ms\,pc^{-3}$)\\
 (1)&  (2) & (3) & (4)  && (5) &(6) &&(7) & (8)   && (9) & (10)&(11)&(12)\\
\hline
      &Core&$0$&$0$&&1.0--3.5&$+0.04\pm0.21$&&$0.3\pm0.1$&$0.7\pm0.2$&&$0.9\pm0.2$&$0.9\pm0.2$&$164\pm30$&$293\pm54$\\
Haf\,9&Halo&$3\pm1$&$12\pm5$&&1.0--4.1&$+1.35\pm0.14$&&$1.6\pm0.2$ &$3.0\pm0.4$&&$21\pm15$&$8.7\pm2.9$&$13\pm4$&$2.1\pm0.7$\\
   &Overall&$3\pm1$&$12\pm5$&&1.0--4.1&$+1.29\pm0.15$&&$1.6\pm0.2$ &$3.1\pm0.5$&&$20\pm14$&$8.5\pm2.7$&$12\pm4$&$2.0\pm0.6$\\
\hline
     &Core&$6\pm2$&$13\pm4$    &&1.7--2.0&$-4.34\pm0.25$&&$0.4\pm0.1$&$0.7\pm0.1$&&$0.7\pm0.1$&$1.2\pm0.2$&$32\pm5$&$22\pm3$\\
Haf\,11&Halo&$52\pm20$&$28\pm9$&&1.7--2.1&$-3.25\pm0.96$&&$1.3\pm0.7$&$2.6\pm1.4$&&$2.1\pm0.9$&$3.6\pm1.6$&$0.8\pm0.3$&$0.05\pm0.02$\\
    &Overall&$56\pm20$&$28\pm9$&&1.7--2.1&$-3.61\pm0.82$&&$1.3\pm0.7$&$2.7\pm1.3$&&$2.0\pm0.8$&$3.6\pm1.4$&$0.8\pm0.3$&$0.05\pm0.02$\\
\hline
       &Core&$1\pm1$&$2\pm2$ &&1.5--2.0&$-3.30\pm0.22$&&$0.3\pm0.1$&$0.5\pm0.1$&&$0.5\pm0.1$&$0.8\pm0.1$&$38\pm5$&$35\pm5$\\
Cz\,32&Halo&$11\pm2$&$22\pm4$&&1.5--2.0&$-1.25\pm0.74$&&$0.9\pm0.5$&$1.6\pm0.8$&&$2.9\pm0.9$&$3.3\pm1.1$&$1.7\pm0.6$&$0.16\pm0.05$\\
   &Overall&$11\pm2$&$22\pm4$&&1.5--2.0&$-1.00\pm0.70$&&$1.0\pm0.5$&$1.7\pm0.8$&&$3.6\pm1.2$&$3.8\pm1.1$&$2.0\pm0.6$&$0.19\pm0.06$\\
\hline
     &Core&$0$&$0$             &&1.1--2.3&$+0.32\pm0.40$&&$0.3\pm0.1$&$0.6\pm0.2$&&$1.8\pm0.8$&$1.0\pm0.2$&$116\pm28$&$165\pm40$\\
Cz\,31&Halo&$7\pm2$&$20\pm6$   &&1.1--2.9&$+1.73\pm0.33$&&$0.7\pm0.2$&$1.2\pm0.3$&&$11\pm8.2$&$4.1\pm1.6$&$8.2\pm3.1$&$1.5\pm0.6$\\
      &Overall&$7\pm2$&$20\pm6$&&1.1--2.9&$+1.48\pm0.31$&&$0.8\pm0.2$&$1.3\pm0.3$&&$11\pm7.9$&$4.1\pm1.5$&$8.1\pm3.0$&$1.5\pm0.5$\\
\hline
       &Core&$2\pm1$&$6\pm3$  &&1.3--2.8&$-0.91\pm0.37$&&$0.3\pm0.1$&$0.7\pm0.2$&&$0.5\pm0.1$&$0.8\pm0.2$&$72\pm23$&$91\pm29$\\
Tr\,13&Halo&$23\pm7$&$71\pm22$&&1.2--3.1&$+1.93\pm0.23$&&$2.9\pm0.5$&$1.4\pm5.3$&&$58\pm43$&$22\pm8.1$&$14\pm5$&$1.4\pm0.5$\\
   &Overall&$24\pm8$&$74\pm25$&&1.2--3.1&$+1.83\pm0.23$&&$3.0\pm0.5$&$1.2\pm1.0$&&$58\pm43$&$29\pm8.1$&$18\pm5$&$1.9\pm0.5$\\
\hline\hline
\end{tabular}
\begin{list}{Table Notes.}
\item The number and mass of the evolved stars in cols.~3 and 4, respectively were derived from
the LFs. Col.~5: MS mass range over which the MF was fitted. Col.~6: MF slope in the MS.
\end{list}
\end{table*}

\subsection{Considerations on dynamical states}
\label{dyna}

Mass segregation in a star cluster scales with the relaxation time, defined as $\rm 
\tr=\frac{N^*}{8\ln N^*}\tcr$, where $\rm\tcr=R/\sigma_v$ is the crossing time, $\rm N^*$ is the 
total number of stars (col.~9 of Table~\ref{tab3}) and $\rm\sigma_v$\ is the velocity dispersion 
(Binney \& Tremaine \cite{BinTre1987}). $\tr$ is the characteristic time scale in which a cluster 
reaches some level of kinetic energy equipartition with massive stars sinking to the core and 
low-mass stars being transferred to the halo. We assume a typical velocity dispersion of 
$\rm\sigma_v\approx3\,\kms$ (Binney \& Merrifield \cite{Binney1998}). Core and overall relaxation 
times are given in cols.~3 and 5 of Table~\ref{tab4}, respectively. However, since the dynamical 
evolution depends strongly on age and cluster mass (Bonatto \& Bica \cite{BB2005}, and references 
therein), it is better characterized by the dynamical-evolution parameter $\rm\tau=age/\tr$. The 
presence of mass segregation and consequently some degree of MF slope flattening in the cores of 
the clusters in the 3Q isample consistent with the large values of $\rm\tau(core)$ (Bonatto \& Bica 
\cite{BB2005}). 

Considering the slope uncertainties, the overall MFs of Haf\,9 and Cz\,31 are similar to a 
standard Salpeter (\cite{Salp55}) IMF ($\chi=1.35$), while the MF of Tr\,13 is steeper. The 
flat MFs of Haf\,11 and Cz\,32 may be affected by the restricted observed MS mass range
($\rm\Delta m_{MS}\approx0.3\,\ms$ and $0.5\,\ms$, respectively) owing to the large distances 
from the Sun (5.2\,kpc and 4.0\,kpc, respectively). In all cases the core MF resulted significantly 
flatter than that in the halo, which reflects the effects of large-scale mass segregation. All 
these pieces of evidence point to advanced dynamical states, especially in the cores of the 3Q 
sample clusters. 

\begin{table}
\caption[]{Dynamical evolution parameters}
\label{tab4}
\renewcommand{\tabcolsep}{1.8mm}
\begin{tabular}{lccccc}
\hline\hline
& \multicolumn{2}{c}{Core} && \multicolumn{2}{c}{Overall} \\
      \cline{2-3}\cline{5-6}\\
Object&\tr&$\tau$&&\tr&$\tau$\\
      &(Myr)&    && (Myr) \\
~~(1)&(2)&(3)&&(4)&(5)\\
\hline
Haf\,9 &$0.34\pm0.10$&$407\pm129$&&$550\pm31$&$2.8\pm1.2$ \\
Haf\,11&$0.71\pm0.23$&$1250\pm453$&&$18\pm6$&$49\pm19$ \\
Cz\,32 &$0.42\pm0.12$&$2660\pm921$&&$19\pm5$&$58\pm20$ \\
Cz\,31 &$0.73\pm0.37$&$248\pm130$&&$25\pm16$& $7.3\pm4.8$   \\
Tr\,13 &$0.30\pm0.10$&$1060\pm390$&&$189\pm126$&$1.7\pm1.1$ \\
\hline\hline
\end{tabular}
\begin{list}{Table Notes.}
\item The dynamical-evolution parameter in cols.~3 and 5 is defined as $\rm\tau=age/\tr$.
\end{list}
\end{table}

\section{Comparison with open clusters in different dynamical-evolution states}
\label{comp}

We compare the 3Q sample clusters with the 631 WEBDA open clusters with parameters (Sect.~\ref{intro}) 
in terms of Galactic longitude and latitude, distance from the Sun and age. A similar analysis of 
13 open clusters studied by means of integrated spectra, using reddening and age histograms 
was carried out by Ahumada et al. (\cite{Ahumada01}).

\begin{figure} 
\resizebox{\hsize}{!}{\includegraphics{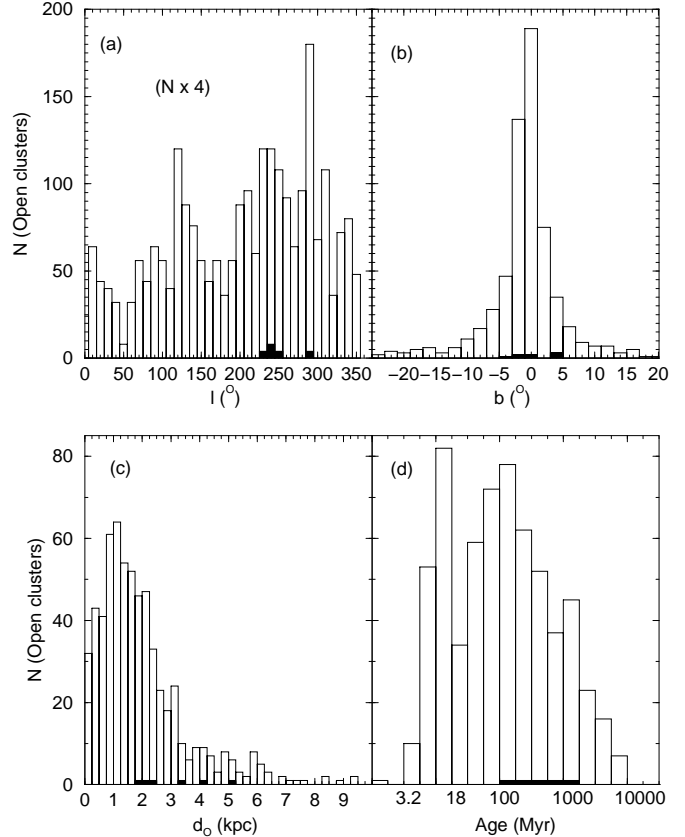}}
\caption[]{Distribution of WEBDA clusters in Galactic longitude (panel (a)),
latitude (panel (b)), distance from the Sun (panel (c)) and cluster age (panel (d)). Filled
histograms: 3Q sample clusters. For clarity the number of clusters in panel (a) was multiplied by 4.}
\label{fig10}
\end{figure}

Galactic longitude and latitude distributions of the WEBDA sample are shown in panels 
(a) and (b) of Fig.~\ref{fig10}, respectively. The clusters are rather uniformly distributed 
in $\ell$ with a slight excess towards the third quadrant, 
where the present clusters are located. This excess probably reflects an observational
bias in the sense that low-contrast clusters are more easily detected towards regions
avoiding the Galactic center and bulge. Alternatively, the excess may reflect a higher 
probability of cluster dissolution towards the Galactic center (e.g. Bergond, Leon, \& 
Guibert \cite{Bergond2001}). In Galactic latitude the majority of the WEBDA
clusters are tightly projected close the Galactic disk, as does also the 3Q sample.

The distribution of the WEBDA open clusters with known distance from the Sun is shown 
in panel (c) of Fig.~\ref{fig10}, where the distribution peaks at
$\rm\ds\approx1.15\,kpc$. Differently from the WEBDA sample, the 3Q sample clusters 
distribute rather uniformly in the range $\rm\ds\approx1.9\,kpc$ to $\rm\ds\approx5.2\,kpc$.

The age distribution of the WEBDA open clusters (panel (d)) presents 2 peaks at $\approx10$\,Myr 
and $\approx100$\,Myr. The 3Q sample clusters distribute rather uniformly in age.

We conclude that the 3Q sample is not atypical compared to the WEBDA sample.

The parameters derived in the previous sections are used to check how the 3Q sample clusters
fit in the context of nearby open clusters characterized by 
different ages and dynamical states. Bonatto \& Bica (\cite{BB2005}) presented a systematic 
analysis of a set of 11 nearby open clusters with distances from the Sun in the range 0.4 -- 1.7\,kpc,
ages from 70\,Myr to 7\,Gyr and total masses from 400 to 5\,300\,\ms. These nearby clusters are M\,26 
(NGC\,6694), NGC\,2516, NGC\,2287 (M\,41), M\,48 (NGC\,2548), M\,93 (NGC\,2447), NGC\,5822, NGC\,2477, 
NGC\,3680, IC\,4651, M\,67 (NGC\,2682) and NGC\,188. The relatively high-Galactic latitude and
rather populous nature of these clusters make them potential sources of accurate intrinsic open cluster 
parameters. As a result, a set of uniform parameters related to the structure (core and overall 
radii, mass and density), dynamical state (core and overall MF slopes, dynamical-evolution parameter 
$\rm\tau=age/\tr$), as well as age and Galactocentric distance of open clusters was obtained. 
Some correlations among these parameters were verified, and a separation of massive 
($\rm m\geq1\,000\,\ms$) and less-massive ($\rm m\leq1\,000\,\ms$) clusters was observed
in some diagrams. The methodology used to derive parameters of the 3Q sample clusters follows 
the same lines as that used in Bonatto \& Bica (\cite{BB2005}).

\begin{figure} 
\resizebox{\hsize}{!}{\includegraphics{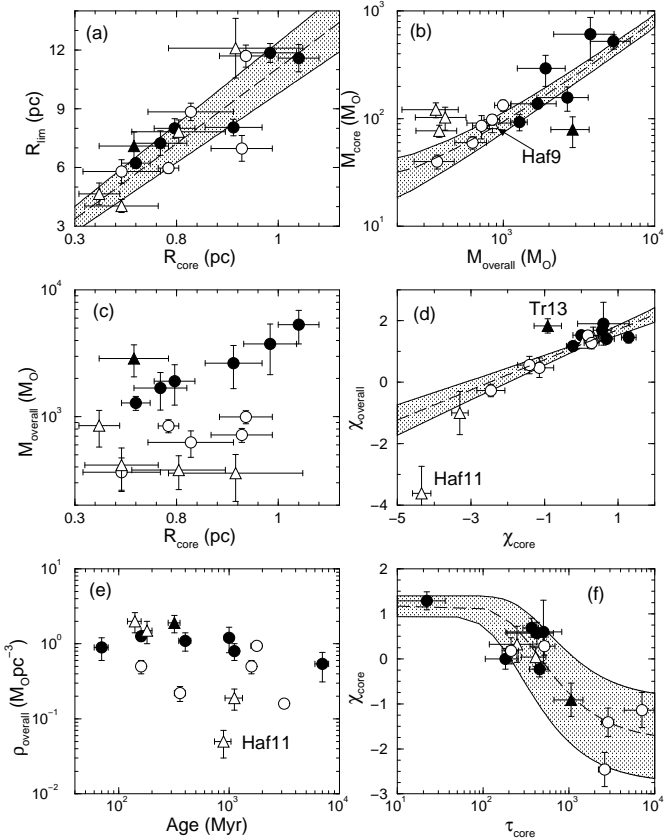}}
\caption[]{Relations involving structural and dynamical-evolution parameters of open clusters. 
Circles: nearby open clusters. Triangles: 3Q sample. Filled symbols: clusters more massive than 
1\,000\,\ms. Open symbols, $\rm m<1\,000\,\ms$. Dashed lines: least-squares fits to the nearby 
clusters (see text). Shaded areas: $\rm1\sigma$ borders of the least-squares fits.}
\label{fig11}
\end{figure}

In Fig.~\ref{fig11} we compare the 3Q sample clusters with those in the nearby cluster set in terms of 
structural and dynamical-evolution parameters. We include in panels (a), (b), (d) and (f) 
the least-squares fits (and the $\rm1\sigma$ borders) of parameter relations derived for 
the nearby clusters in Bonatto \& Bica (\cite{BB2005}). 

Within uncertainties the 3Q sample clusters fit well in the relation involving core and limiting 
radii (panel (a)). In this case the least-squares fit is given by $\rl=(1.05\pm0.45)+
(7.73\pm0.66)\times\rc$, with a correlation coefficient $\rm CC=0.95$, using the reference 
sample. 

In the relation involving core and overall mass (panel (b)) the 3Q sample clusters fall slightly
off the $\rm1\sigma$ border of the comparison sample relation, except for Haf\,9 which fits 
in it. The least-squares fit is $\rm M_{core}=(14.17\pm10.02)+(0.08\pm0.01)\times M_{overall}$, 
with $\rm CC=0.92$. 

Within the uncertainties the separation observed among the massive and less-massive nearby 
clusters (panel (c)) can be extended to the 3Q sample clusters. 

In the $\rm\chi_{core}\times\chi_{overall}$ plot (panel (d)) Tr\,13, and especially Haf\,11,
deviate from the relation. In this case the least-squares fit is 
$\rm\chi_{overall}=(1.22\pm0.07)+(0.49\pm0.08)\times\chi_{core}$, with $\rm CC=0.90$.

Compared to the nearby clusters of the same age range Haf\,11 seems to present an exceedingly low 
overall density (panel (e)), which might indicate an underestimation of the overall mass. This 
effect may be accounted for observationally by the relatively large distance from the Sun and low 
latitude of Haf\,11 (Sect.~\ref{dyna}). 

\begin{figure} 
\resizebox{\hsize}{!}{\includegraphics{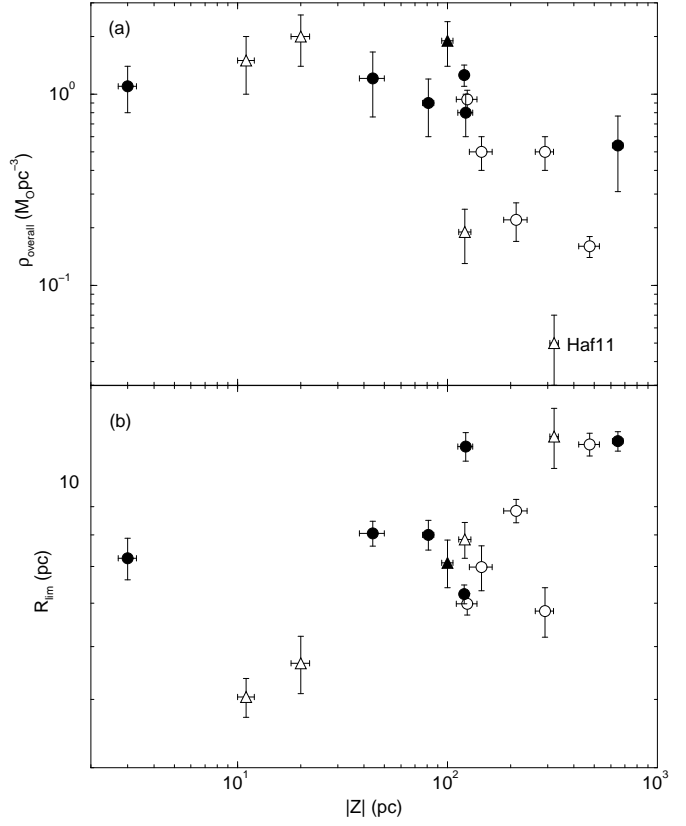}}
\caption[]{Relations of overall density (top panel) and limiting radius (bottom panel) 
with the distance to the Galactic plane. Symbols as in Fig.~\ref{fig12}.}
\label{fig12}
\end{figure}

Finally, with respect to the core MF slope vs. dynamical-evolution parameter (panel (f)) the 
least-squares fit is $\rm\chi_{core}=(1.17\pm0.23)-(3.0\pm0.7)
\exp{-(\frac{439\pm156}{\tau_{core}})}$, with $\rm CC=0.82$. Compared to the nearby cluster set, 
the cluster cores in the 3Q sample present evidence of important dynamical evolution, such as flat 
MF slopes and large values of $\rm\tau_{core}$. This happens despite the young age of Haf\,9 
($\rm\sim140\,Myr$), Cz\,31 ($\rm\sim180\,Myr$) and Tr\,13 ($\rm\sim320\,Myr$) as compared to the 
nearby open cluster sample. A possible explanation for this evidence of advanced dynamical evolution 
for open clusters with ages in the range 100 to 400\,Myr may be orbital planes closer to the 
Galactic plane.

The dependence of limiting radius on distance to the plane ($\rm|z|$) occurs for the less-massive
clusters, both for the nearby clusters and those in the 3Q sample (bottom panel of Fig.~\ref{fig12}). 
For the massive clusters the dependence of \rl\ on $\rm|z|$ is less important. In the top panel of 
Fig.~\ref{fig12} we show the relation of the overall density with $\rm|z|$. A correlation shows up
for the less-massive clusters in the sense that higher-$\rm|z|$ clusters have lower overall densities.
However, we note that the limiting radius may be underestimated (Sect.~\ref{struc}). The plots in 
Fig.~\ref{fig11} (and other parameter relations in Bonatto \& Bica \cite{BB2005}) appear to be useful 
tools in the interpretation of the structure and dynamical states of open clusters.

\section{Concluding remarks}
\label{Conclu}

In this paper we analyzed 5 low-contrast (observed contrast parameter $\rm\delta^\prime_c=1.9 - 4.7$) 
open clusters located in the third quadrant by means of 2MASS photometry. The clusters did not 
have available parameters in the literature. The present approach involves Colour-Magnitude 
Diagrams, statistical field-star subtraction, colour-magnitude filters, radial density profiles 
and luminosity/mass functions built for the core, halo and overall cluster regions. In all cases
the CMD morphology resulting from the field-star statistical subtraction is unambiguously 
characteristic of open clusters of different ages with a well-defined main sequence, turnoff and, 
for Haf\,11 and Cz\,32, the presence of a conspicuous giant clump.

We derived reddening values \ebv\ in the range 0.06 to 0.66, ages from 140\,Myr to 1.1\,Gyr, 
distances from the Sun from 1.9 to 5.2\,kpc, and total masses from 360 to 2\,900\,\ms. 4 out 
of the 5 clusters are located up to $\sim3.4$\,kpc outside the Solar circle. 

By applying the colour-magnitude filter the two-parameter King profile could be fitted 
to the radial density distribution of all clusters producing core radii in the range 
$\rm0.42\leq\rc(pc)\leq 1.09$ and limiting radii in the range $\rm4.0\leq\rl(pc)\leq 12.1$. The 
resulting contrast parameters (after applying the colour-magnitude filter) increase to the range 
$\rm\delta_c=6.4-11.2$, about 3 times as large as the observed ones.

In all cases the clusters in the 3Q sample present signs of advanced dynamical evolution, such 
as flat core mass function slopes, large values of the dynamical-evolution parameter, and 
large-scale mass segregation. This effect occurs despite the relatively young cluster ages 
in part of the sample (140 to 320\,Myr), resembling dynamical states of the nearby Gyr-old 
clusters (Bonatto \& Bica \cite{BB2005}). Proximity to the Galactic plane may have accelerated 
their dynamical evolution.

We conclude that 2MASS photometry can be used to study the large number of unexplored 
low-contrast open clusters when associated with statistical field-star decontamination  
(for CMD morphology and isochrone fit) and colour-magnitude filters (for structure and
mass-function analyses).

\begin{acknowledgements}
We thank the anonymous referee for important remarks which improved the paper.
This publication makes use of data products from the Two Micron All Sky Survey, which 
is a joint project of the University of Massachusetts and the Infrared Processing and 
Analysis Center/California Institute of Technology, funded by the National Aeronautics 
and Space Administration and the National Science Foundation. We also made use of the 
WEBDA open cluster database. We acknowledge support from the Brazilian Institution CNPq.
\end{acknowledgements}

%

\end{document}